\documentclass[cits]{Pos}
\def\kaos{{\sc Kaos}\@}
\graphicspath{{figures/}}

\title{Current Issues in Kaon Photoelectro-Production off the Nucleon}

\ShortTitle{Current Issues in Kaon Photoelectro-Production off the Nucleon}

\author{\speaker{Patrick ACHENBACH}
  \\
  Institut f\"ur Kernphysik, J. J. Becherweg 45,\\
  Johannes Gutenberg-Universit\"at, D-55099 Mainz, Germany\\
  E-mail: \email{patrick@kph.uni-mainz.de}}

\abstract{The electromagnetic kaon production amplitudes associated to
  $\Lambda/\Sigma^0$ hyperons can be described by phenomenological
  models, most notably by isobar approaches.
 
  Experimental data on kaon production have been collected at ELSA,
  SPring8, GRAAL, LNS Tohoku, and Jefferson Lab in the past, the
  measurements at Jefferson Lab providing the largest kinematic
  coverage and statistical significance. However, ambiguities inherent
  in the models, some data inconsistency in the cross-sections taken
  at different laboratories, and the problem of missing acceptance in
  forward direction of the experimental set-ups hinders a reliable
  extraction of resonance parameters.

  Predictions for the hypernuclear photo-production cross-section rely
  on a consistent and comprehensive description of the elementary
  process at forward kaon angles, where the current strong variation
  of the models is very unsatisfactory.

  A number of new experiments are now addressing these issues, among
  them the charged kaon electro-production programme with the \kaos\
  spectrometer at the Mainz Microtron MAMI.  In this work predictions
  of the two prominent isobar models, Kaon-Maid and Saclay-Lyon A, are
  compared for the kinematics at MAMI.  }

\FullConference{XLVIII International Winter Meeting on Nuclear Physics
  in Memoriam of Ileana Iori \\
  25-29 January 2010\\
  Bormio, Italy}

\begin{document}

% -------------------------------------------------------------------
\section{Introduction}
% -------------------------------------------------------------------
% 
The kaon electro-production reaction on a proton target is a promising
channel for studying the excited nucleon. During the last decade, key
kaon production measurements were performed at ELSA, the Electron
Stretcher Accelerator in Bonn, published by the SAPHIR
collaboration~\cite{Tran1998,Glander2003}, and at Jefferson Lab
(Thomas Jefferson National Accelerator Facility).  Experiment E98-108
in Hall~A took data in a kinematic region of high
squared-momentum-transfer $Q^2= 2.35$\,(GeV$/\!c$)$^2$ and hadronic
energy $W \ge$ 1.80~\cite{Iodice2003}.  Several large-acceptance
measurements have been performed with the CLAS detector at
Hall~B~\cite{Carman2003,Ambrozewicz2007}.  New measurements at CLAS
have been performed with improved statistical significance and wider
energy coverage compared to previous results~\cite{McCracken2010}.
However, the detector CLAS has a limited acceptance for four-vector
momentum transfers $Q^2< 0.5$\,(GeV$/\!c$)$^2$ and for forward angles.
Experiment E03-018 carried out at Hall~C studied the two hyperon
channels p$(e,e'K^+)\Lambda$ and p$(e,e'K^+)\Sigma^0$ and performed
the first precise Rosenbluth separation of cross-sections into
longitudinal and transverse terms~\cite{Niculescu1998,Mohring2003}.
In spite of all these data, cross-sections published after 1990 and
taken at different laboratories are inconsistent and most of the
experimental set-ups used to study the strangeness production channels
are missing acceptance in forward acceptance.

Although it is believed that the Standard Model provides the right
theoretical basis for the description of the process, the
non-perturbative nature of QCD at low energies precludes a direct
comparison of experiments and theory.  In analogy to the successful
description of pion photoproduction in the $\Delta$-resonance region
or $\eta$ photoproduction in the second resonance region, the
electromagnetic kaon production amplitudes associated to
$\Lambda/\Sigma^0$ hyperons can be described by phenomenological
models.  Theoretical groups have developed a particular type of
effective Lagrangian model, commonly referred to as isobar approach,
in which the reaction amplitude is constructed from $s$-, $t$-, or
$u$-channel exchange diagrams.  Most of the models use single-channel
approaches, in which a single hadron is exchanged.  Since several
resonances may contribute in this channel, models disagree on their
relative importance, and many free parameters have to be fixed.
 
There are still a number of open problems in the interpretation of
kaon photoelectro-production data and the description of the process
using phenomenological models, see Ref.~\cite{Mart2009} for a recent
discussion.  Especially the shape discrepancy at $W \approx$ 1.9\,GeV
is problematic, and partial-wave analyses in this energy region have
produced various resonance contributions, including $D_{13}$,
$P_{13}$, $D_{11}$, and $S_{11}$ states.

To conclude, it is fair to say that new experimental data on
strangeness production will challenge and improve our understanding of
the strong interaction in the low energy regime of QCD.

% -------------------------------------------------------------------
\section{Phenomenological Description of the Reaction}
\begin{figure}
  \centering
  \includegraphics[width=0.9\textwidth]{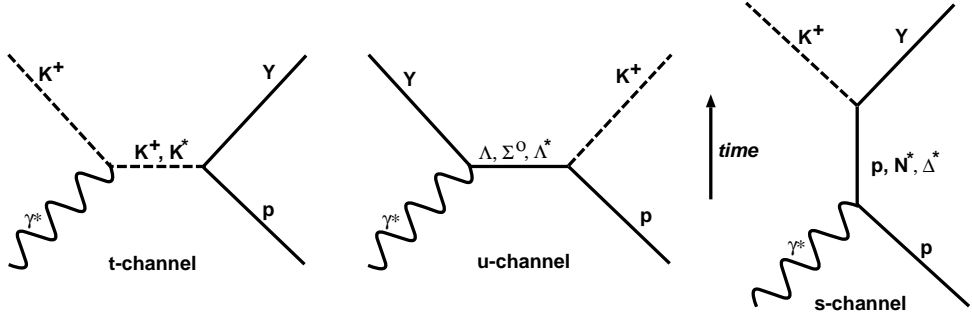}
  \caption{Lowest-order Feynman diagrams of $s$-, $t$-, and
    $u$-channel exchanges for the p$(e,e'K^+)Y$ reaction. The vertices
    enter into the models through coupling constants and
    form-factors. The different resonance contributions are subject to
    model variations and global fits.}
  \label{fig:feynman}
\end{figure}

% -------------------------------------------------------------------
% 
In isobaric models the reaction amplitudes are constructed from
lowest-order (so-called tree-level) Born terms with the addition of
extended Born terms for intermediate particles, $N$, $K$, or $Y$
resonances, exchanged in the $s$-, $t$-, and $u$-channels as shown in
Fig.~\ref{fig:feynman}. Each intermediate state enters into the model
through its coupling constant and decay width. Diagrams containing
intermediate nucleon resonances can produce peaks at given hadronic
energies $W$, or equivalently $\sqrt{s}$, in the cross-sections at the
pole masses of the $s$-channel intermediate states. Intermediate kaons
and hyperons cause no such peaking behaviour, and are often called
background contributions.

A complete description of the reaction process would require all
possible channels that could couple to the initial and final
state. Most of the model calculations for kaon electrophoto-production
have been performed in the framework of tree-level isobar
models~\cite{Adelseck1990,Williams1992,Bennhold1999}, however, only
few coupled-channels calculations exist~\cite{Feuster1999,Shyam2010}.
In the tree-level framework higher-order mechanisms like final-state
interactions and channel couplings are not treated explicitly.

The advantages of the one-channel, tree-level approach are its limited
complexity and the identification of the dominant trends.  Even so,
several dozen parameters remain. One reason is that, in contrast to
pion and eta production, the kaon production process in the
$s$-channel is not dominated by a single resonant state.  Although the
choice of the resonances is guided by existing $\pi N$ data and quark
model predictions, the models differ in the use of specific nucleon,
hyperon, and kaon resonances.  The drawback of the isobaric models is
the large and unknown number of exchanged hadrons that can contribute
in the intermediate state of the reaction. Depending on which set of
resonances is included, very different conclusions about the strengths
of the contributing diagrams for resonant baryon formation and kaon
exchange may be reached.

For an unpolarised electron beam and an unpolarised target, the
five-fold differential cross-section for the p$(e,e'K^+)\Lambda$
process can be written, see {\em e.g.\/}~\cite{Amaldi1979}, in a very
intuitive form:
\begin{equation}
   \frac{d\sigma}{d E_{e'} d\Omega_e d\Omega_K^*}
   = \Gamma_v \frac{d \sigma_v}{d\Omega_K^*}\,,
\label{eq:xsec}
\end{equation}
where the virtual photo-production cross-section is conventionally
expressed as
\begin{equation}
   \frac{d\sigma_v}{d\Omega_K^*}
   = \frac{d\sigma_T}{d\Omega_K^*}
   + \epsilon \frac{d\sigma_L}{d\Omega_K^*}
   + \sqrt{2 \epsilon (1 + \epsilon)} \,
     \frac{d\sigma_{LT}}{d\Omega_K^*} \cos\phi_K^*
   + \epsilon \frac{d\sigma_{TT}}{d\Omega_K^*} \cos 2 \phi_K^*\, .
\label{eq:diffxsec}
\end{equation}
The kaon angles $\theta_K^*$ and $\phi_K^*$ are given in spherical
coordinates in the hadronic centre-of-mass system.  The
degree-of-polarisation of the virtual photon is denoted by
$\epsilon$. The terms indexed by $T, L, LT, TT$ are the transverse,
longitudinal and interference cross-sections. The electro-production
of strangeness introduces two contributions, that are vanishing for
the kinematic point at $Q^2 =$ 0: the longitudinal coupling of the
photons in the initial state, and the electromagnetic and hadronic
form factors of the exchanged particles. The introduction of form
factors has a serious impact on the properties of a model.  It is
common practice to use phenomenological form factors to account for
the extension of the point-like interactions at the hadronic
vertices~\cite{Janssen2003}. It is a well-known phenomenon that the
insertion of strong form factors breaks the gauge invariance of isobar
models for meson photo- and electro-production. Different models
typically have different prescriptions for restoring gauge invariance.

% -------------------------------------------------------------------
\section{Kaon-Electroproduction at MAMI}
% -------------------------------------------------------------------
%
A first experiment of $\Lambda$ and $\Sigma^0$ hyperons in elementary
electro-production at MAMI was carried out at the spectrometer
facility of the A1 Collaboration at the Institut f\"ur Kernphysik in
Mainz, Germany. During the last years the facility has been extended
by the magnetic spectrometer \kaos, dedicated to the detection of
charged kaons. The electron beam impinged with an energy of 1507\,MeV
on a liquid-hydrogen target.  The data were taken at two different
settings with kaons in a large in-plane angular range, $\vartheta_K=$
21--43$^\circ$, and in the momentum range of 400--700\,MeV$/c$. The
two settings were at four-vector momentum transfers of $\langle
Q^2\rangle =$ 0.050\,(GeV$/c$)$^2$, respectively 0.036\,(GeV$/c$)$^2$,
and at hadronic energies of $\langle W\rangle =$ 1.67\,GeV,
respectively 1.75\,GeV.  Cross-sections are to be extracted in the
near future.  The detection of kaons at very forward laboratory angles
will be achieved in the near future by steering the primary beam
through the spectrometer and a magnetic chicane comprising two
compensating sector magnets.

\begin{figure}
  \centering
  \includegraphics[angle=90,width=0.49\textwidth]{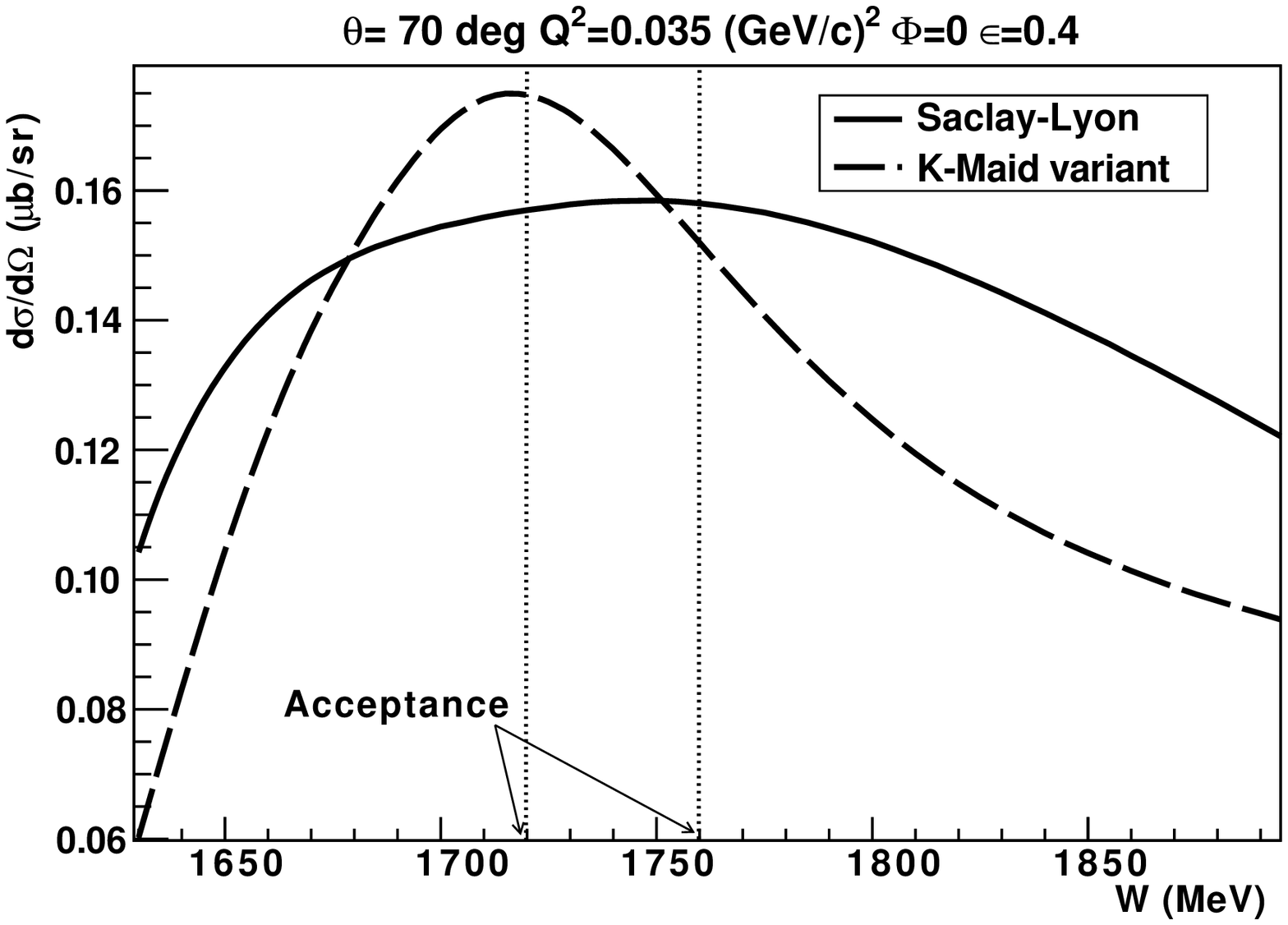}\hfill
  \includegraphics[angle=90,width=0.47\textwidth]{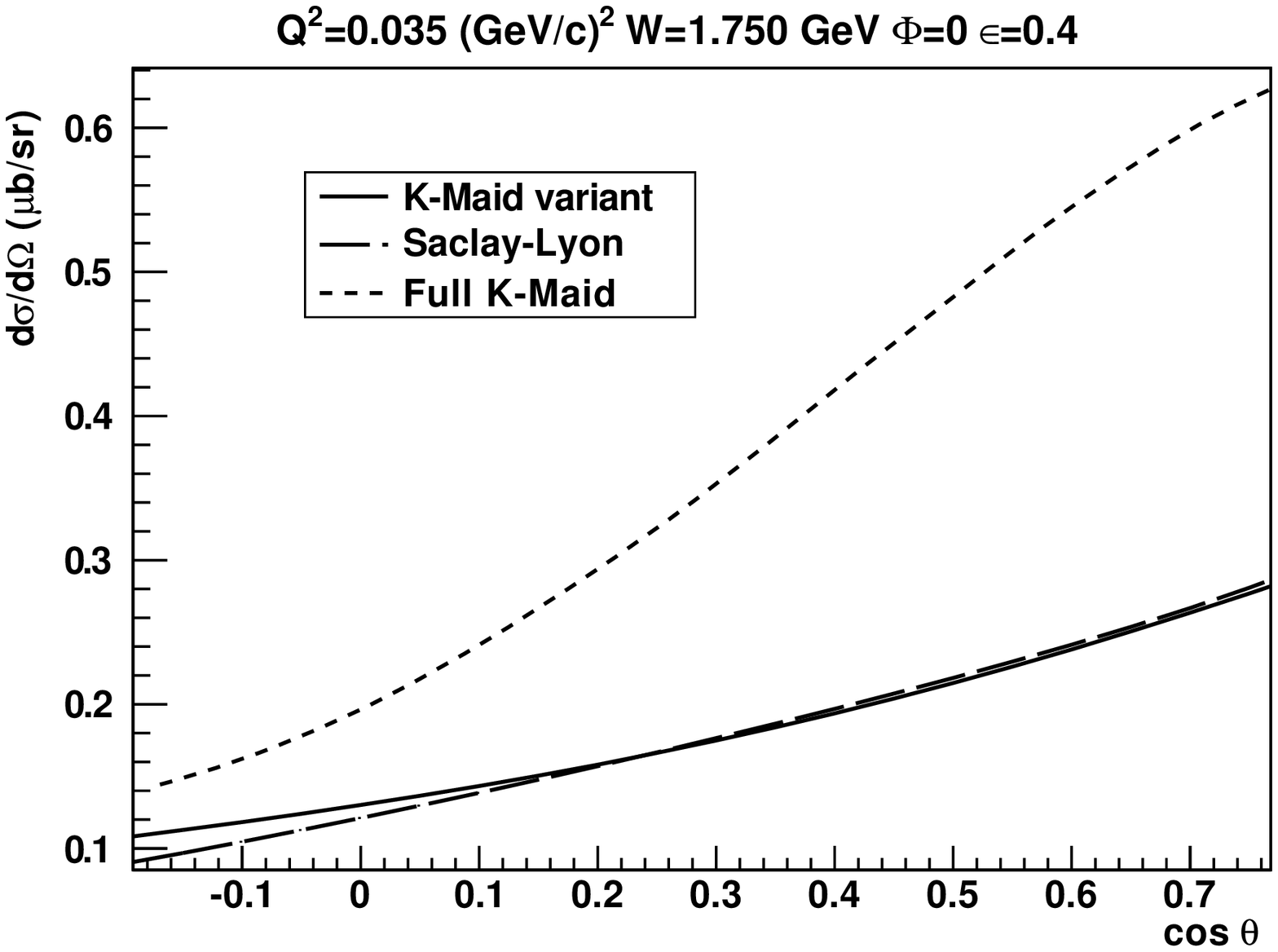}
  \caption{Comparison of cross-section predictions from the Kaon-Maid
    and Saclay-Lyon models for the kinematics measured at MAMI as a
    function of the hadronic energy (left) and as a function of the
    centre-of-mass kaon scattering angle (right). The hadronic energy
    acceptance of the spectrometer set-up at MAMI for the second
    kinematic setting is indicated. The acceptance in $\cos \theta$ is
    between 0.1 and 0.7.}
  \label{fig:KMaidSaclayLyon}
\end{figure}

Model predictions for centre-of-mass cross-sections as a function of
the hadronic energy at fixed kaon angle and for the differential
cross-section at fixed hadronic energy are shown in
Fig.~\ref{fig:KMaidSaclayLyon}. The predictions are from the Kaon-Maid
model, a variant of it, and the Saclay Lyon~A (SLA)
model~\cite{Mizutani1998}.  Common to the Kaon-Maid and Saclay-Lyon
models is that, besides the extended Born diagrams, they also include
kaon resonances $K^*(890)$ and $K_1(1270)$.  In Kaon-Maid, four
nucleon resonances, the $S_{11}(1650)$, $P_{11}(1710)$,
$P_{13}(1720)$, and the ``missing resonance'' $D_{13}(1900)$ have been
included. This $D_{13}$ state has never been observed in pionic
reactions but the existence of this resonance with considerable
branching into the strange channel was predicted by the constituent
quark model calculations.

\begin{figure}
  \centering
  \includegraphics[width=0.49\textwidth]{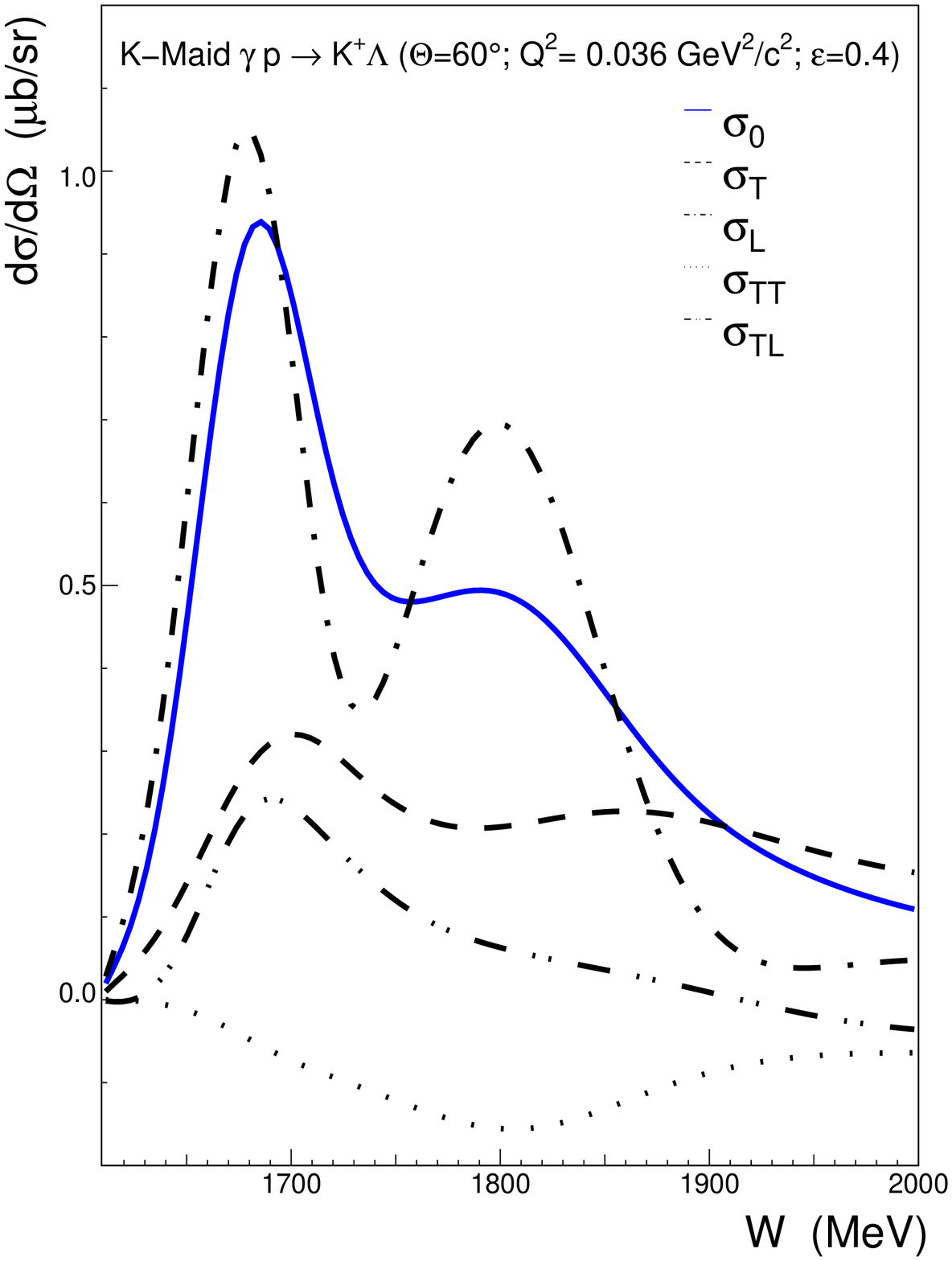}
  \includegraphics[width=0.49\textwidth]{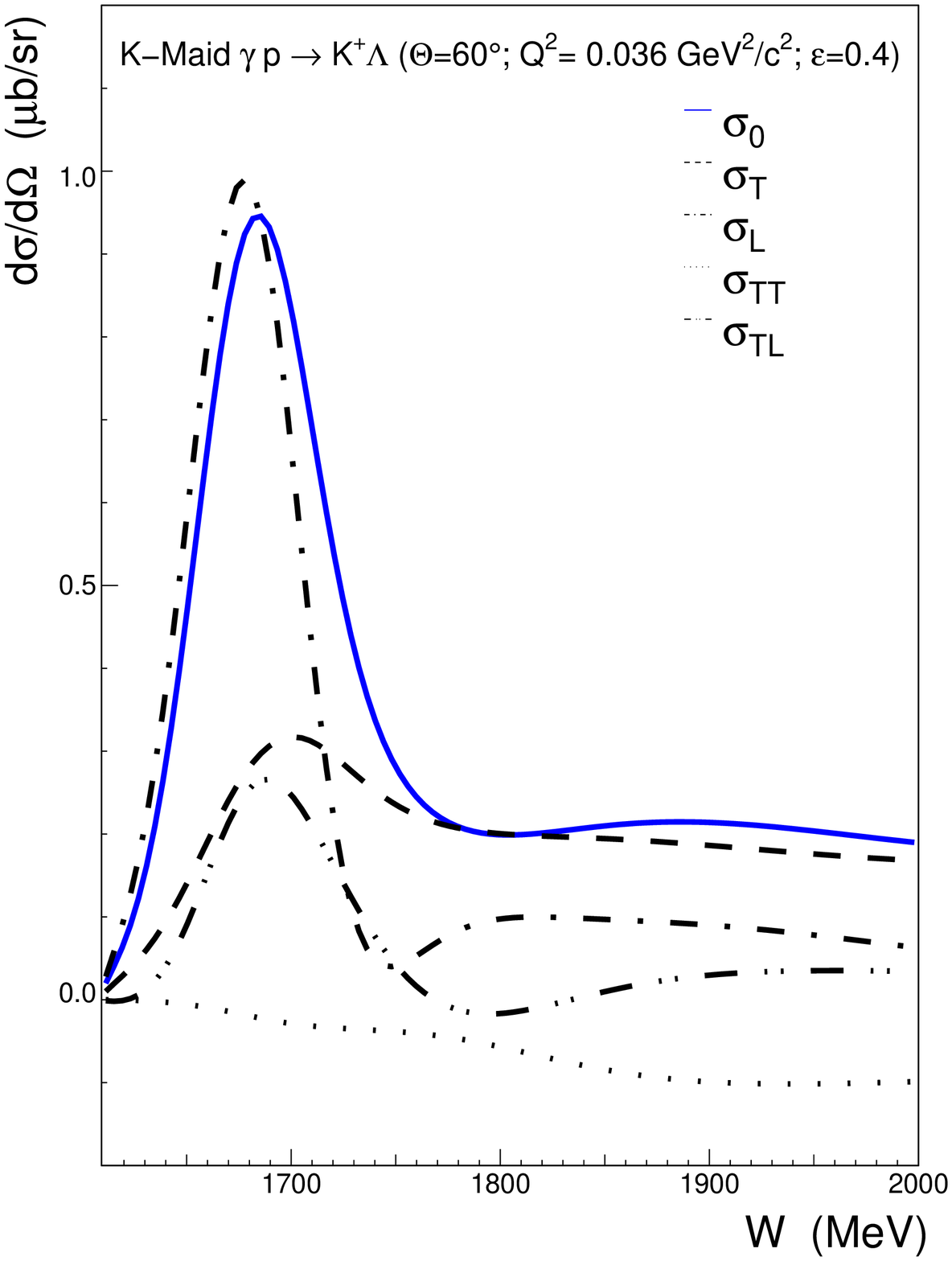}
  \caption{Kaon-Maid cross-section predictions for the $\Lambda$
    reaction channel in the kinematics measured at MAMI using the full
    version (left) and a variant without longitudinal coupling of the
    $D_{13}(1900)$ resonance (right). The definition of the separated
    terms of the cross-section is given in the text.}
  \label{fig:KMaidPredictions}
\end{figure}

Calculations with both models for the kinematics studied with the
\kaos\ spectrometer at MAMI result in relatively large discrepancies
in the longitudinal contributions to the cross-section. The full
Kaon-Maid model including all couplings and individual dipole
form-factors of baryon resonances reveals the following set of
cross-sections at the centre of the acceptance in the kinematics
probed at MAMI: $\sigma_T=$ 235\,nb$/$sr; $\sigma_L=$ 423\,nb$/$sr;
$\sigma_{TT}=$ $-$121\,nb$/$sr, $\sigma_{TL}=$ 120\,nb$/$sr. The large
longitudinal as well as the $\sigma_{TT}$ and $\sigma_{TL}$ cross-sections
are strongly affected by the longitudinal coupling of the
$D_{13}(1900)$ resonance and the appropriate electromagnetic
form-factor.  With vanishing longitudinal coupling the change in
$\sigma_{L}$, $\sigma_{TT}$, and $\sigma_{TL}$ cross-sections is
dramatic. This behaviour of the model is questionable.
Fig.~\ref{fig:KMaidPredictions} shows the predictions using the full
version compared to a variant without longitudinal coupling of the
$D_{13}(1900)$ resonance in the kinematic region of the measurements
at MAMI.  The electromagnetic form-factor of the $D_{13}(1900)$ was
introduced into Kaon-Maid and fixed to the data taken at Jefferson
Lab~\cite{Niculescu1998,Mohring2003} to produce a sharp peak in
$\sigma_L$ as a function of $Q^2$ at one hadronic energy ($W=$
1.84\,GeV).

The model SLA is a simplified version of the full Saclay-Lyon model in
which a nucleon resonance with spin-5$/$2 appears in addition.  The
extension of the Saclay-Lyon model to the electro-production does not
include too many additional parameters. Predictions of the Saclay-Lyon
model are in good agreement with predictions of the discussed variant
of the Kaon-Maid model in which identical phenomenological
electromagnetic form-factors are used and also with those from the
original Kaon-Maid version without the longitudinal coupling of the
$D_{13}(1900)$ resonance.

These models characterise our present understanding of kaon
photo-production reactions at photon energies below 1.5\,GeV. However,
the variations in the model predictions in the resonance region are
clearly visible which lead to the conclusion, that our description of
the process still lacks considerable insight.

% -------------------------------------------------------------------
\section{Perspective}
% -------------------------------------------------------------------
%
With regard to future experimental work it is obvious that there are
interesting topics in elementary kaon electro-production in the region
$0 < Q^2 < 0.5$\,GeV$\!/c^2$.  The planned operation of the \kaos\
spectrometer using a pre-target beam chicane provides an experimental
set-up that is unique in the world and which will allow unprecedented
measurements at forward angles.

Kaon electro-production measurements are only one of the many
perspectives from which strangeness physics should be viewed, with the
ultimate goal of providing a more thorough understanding of the
composition of nuclear matter and sub-nuclear reaction dynamics.

% -------------------------------------------------------------------
\section*{Acknowledgements}
% -------------------------------------------------------------------
%
I am thankful to Terry Mart from the University of Indonesia and Petr
Byd\v{z}ovsk\'{y} from the Nuclear Physics Institute in {\v R}e{\v z}
near Prague for sending me their codes for the calculation of kaon
photoelectro-production cross-sections and who assisted me in the
interpretation of the results.

\end{document}